\documentclass{optica-article}

\journal{opticajournal} 

\articletype{Research Article}

\usepackage{lineno}

\usepackage{xcolor}
\usepackage{siunitx}

\usepackage{hyperref}
\usepackage{soul}
\hypersetup{colorlinks=true,citecolor={blue},linkcolor={blue},urlcolor={blue}}

\begin{document}

\title{Observation of $\sigma$-$\pi$ coupling and mode selection in optically trapped artificial polariton molecules}

\author{Krzysztof~Sawicki,\authormark{1,*} Valtýr~Kári~Daníelsson,\authormark{2} Dmitriy~Dovzhenko,\authormark{1} Pavlos~G.~Lagoudakis,\authormark{3} Simone~De~Liberato,\authormark{1,4} and Helgi~Sigurðsson\authormark{2,5}}

\address{\authormark{1}School of Physics and Astronomy, University of Southampton, Southampton SO17 1BJ, United Kingdom\\
\authormark{2}Science Institute, University of Iceland, Dunhagi 3, IS-107 Reykjavik, Iceland\\
\authormark{3}Department of Physics, Faculty of Applied Mathematical and Physical Sciences, National Technical University of Athens, 15772 Athens, Greece\\
\authormark{4}Istituto di Fotonica e Nanotecnologie, Consiglio Nazionale delle Ricerche (CNR), Piazza Leonardo da Vinci 32, Milano, 20133, Italy\\
\authormark{5}Institute of Experimental Physics, Faculty of Physics, University of Warsaw, ul.~Pasteura 5, PL-02-093 Warsaw, Poland}

\email{\authormark{*}k.sawicki@soton.ac.uk} 


\begin{abstract*} 
Microcavity exciton-polariton condensates under additional transverse confinement constitute a flexible optical platform to study the coupling mechanism between confined nonequilibrium and nonlinear states of matter. Driven far from equilibrium, polariton condensates can display spontaneous synchronization and instabilities depending on excitation and material parameters, showcasing emergent and intricate interference patterns based on mode competition over mutual gain landscapes. Here, we explore this coupling mechanism between polariton condensates populating the first excited $p$-state manifold of coupled optically trapped condensates and show a rich structure of patterns based on excitation parameters. The optical reconfigurability of the laser excitation patterns enables the creation of an annular-shaped beam to confine polaritons in a tailored trapping potential, whilst the dissipative nature of the optical traps enables effective interaction with neighboring condensates. Our results underpin the potential role of polariton condensates in exploring and simulating $\sigma$ and $\pi$ molecular bonding mechanisms between artificial two-dimensional diatomic orbitals and beyond.
\end{abstract*}

\section{Introduction}
Analogue simulation over interatomic interactions can provide crucial insight into complex physics ranging from chemical reactions, buildup and decay of correlations, dynamical processes, etc., that are beyond the reach of classical computing strategies. Quantum computing platforms offer a clear advantage in simulating many-body physics~\cite{Feynman:IntJTheorPhys1982} but still face many challenges in engineering and design~\cite{Fauseweh_Natcomm2024}. Ideally, an analogue simulator should come with many tunable and rewritable parameters to artificially recreate as many target systems as possible, while also permitting easy readout of information. Optically driven and addressable systems such as microdisk lasers~\cite{Fasching_OptExp2009, Parto_Nanophotonics2020}, microspheres~\cite{Rivera2018}, microrings~\cite{Zhang2019}, planar microcavities~\cite{Bayer_PRL1998, Dufferwiel_APL2015, Urbonas_ACSPhot2016, Sawicki:Nanophotonics2021, Dovzhenko:PRB2023, Dovzhenko:arxiv2025} and micropillars~\cite{Vasconcellos_APL2011, Galbiati_PRL2012, Zambon:PRA2020}, and optically trapped polaritons~\cite{Cristofolini_PRL2013, Askitopoulos_PRB2015, Pieczarka_PRB2019, Sawicki:PRB2024} offer a unique pathway to investigate the spatial coupling between high-order artificial photonic atoms. The high-field-seeking character of pumped lasing systems results in phase-locking and synchronization between distinct overlapping degenerate lasing modes that aim to optimize their gain and mutual amplitude. Adapting this idea to molecular chemistry, complex chemical structures can be mapped by optical platforms whose transverse lasing states imitate the binding of atomic orbitals ($s$, $p$, $d$, etc.) into molecular orbitals [see schematic Fig.~\ref{fig1}].

One of the fundamental limitations of optical simulators of quantum systems is the weak interaction of purely photonic systems, which greatly limits the possibilities of examining coupled systems and external control using electric and magnetic fields. Instead, following recent proposals~\cite{Cherotchenko:PRB2021}, we demonstrate a reconfigurable platform based on coupled exciton-polariton condensates in planar microcavities for simulation of artificial two-dimensional molecules~\cite{Johnston_PRB2021}. Cavity exciton-polaritons are bosonic quasiparticles appearing in the strong coupling regime between quantum well excitons and Fabry-Pérot photons. Their physics lies at the interface of condensed matter and optics, making them intriguing candidates to explore the dynamics of quantum fluids optically driven out of equilibrium~\cite{Carusotto_RMP2013}. Their (relatively) strong Coulomb interactions and light photonic effective mass allow them to undergo power-driven condensation into a macroscopically occupied quantum state referred to as nonequilibrium Bose-Einstein condensates. To date, exciton-polariton systems have been used to simulate a number of phenomena~\cite{Kim_springer2017} such as band structures in artificial lattices~\cite{Amo:ComptesRendusPhysique2016}, the XY~\cite{Berloff:NatMat2017, Tao_NatMat2022} and Ising spin Hamiltonian~\cite{Alyatkin_SciAdv2024}, topological insulating materials~\cite{Klembt_Nature2018}, non-Abelian gauge fields and spin-orbit coupling~\cite{Polimena_Optica2021}, photon localization~\cite{Jamadi:Optica2022}, Floquet physics~\cite{Redondo_NatPhot2024}, and more. The sensitivity of polaritons to external fields offers many in-situ tuning possibilities, and the advantage of optically induced potentials in polariton systems is their flexible reconfigurability through spatial light modulation, meaning they do not require a time-consuming, costly, and irreversible sample fabrication process. The emitted cavity light from the polariton molecules allows measurement of amplitude, phase, polarization (pseudospin), frequency, and momentum degrees of freedom using standard optical techniques.

In this work, we demonstrate the application of exciton-polariton (from here on {\it polariton}) condensates as an optically accessible solid-state platform to explore high-order coupled artificial atoms. The nature of this optical system allows for efficient engineering of coupling and exploration of a vast parameter space, which is expected from molecular simulators as well as optical graphs and applications in topological polaritonics. Polaritons are realized in optically pump-induced traps, due to their repulsive interactions with the photoexcited incoherent exciton background. Importantly, polariton condensates can form in high-order transverse modes of the trap when pumped sufficiently close to threshold~\cite{Cristofolini_PRL2013, Dreismann_PNAS2014, Askitopoulos_PRB2015, Sun_PRB2018, Topfer_PRB2020, Sitnik_PRL2022, Pieczarka_OptExpr2022, Aladinskaia_PRB2023, Barrat_SciRep2024, Zaremba_AdvOptMat2025} due to an intricate balance between gain-and-losses~\cite{Nalitov_PRA2019, Bochin_OptMatExpr2023}. In particular, we investigate the hybridization of condensate wavefunctions of $p$-orbital type between two and three traps. Our scheme goes beyond the hybridization of lowest-order ($s$-orbital) condensate wavefunctions demonstrated recently in subwavelength polariton gratings~\cite{Gianfrate_NatPhys2024}. In particular, we present an optical system of artificial polariton molecules, in which we demonstrate the possibility of manipulating the strength and type of coupling, creating systems resembling $\sigma$ and $\pi$ bonds of $p$ orbitals in homoatomic molecules and $s$-$s$, $s$-$p$, $s$-$d$ in heteroatomic molecules. Apart from analogy to binding of artificial atoms, our optical system benefits from flexible write-in and read-out of information about the ``atomic'' degrees-of-freedom with highly reconfigurable strength and type of coupling, enabling switching between different phases and geometry of interaction as well as symmetric and asymmetric coupling between two optical objects with discrete energy states. 

\section{Results}
Our system is a high-quality, strain-compensated GaAs-based planar microcavity with multiple quantum wells containing Wannier-Mott excitons, which couple strongly with confined photons to form exciton-polaritons~\cite{Cilibrizzi:APL2014}. Experiments are performed under continuous-wave, nonresonant optical pumping at cryogenic temperatures, which support the condensation of polaritons around $\approx7$ K, as observed through photoluminescence (PL) collected via a microscope objective. For more details on the sample and the experimental procedure, see Methods.

In order to realize the optical trap for the polariton condensates, we use a spatial light modulator (SLM) to create an annular-shaped pump profile of zero angular momentum. While an axicon setup could also be used to create such a beam profile~\cite{Askitopoulos_PRB2015} the SLM technique allows us to easily create two traps at the same time of adjustable separation distance to explore the coupling mechanism between the condensates within~\cite{Khan_NJP2017, Harrison:PRB2020}. The nonresonant pump photoexcites a co-localized high-energy charge carrier distribution, which undergoes fast relaxation down in energy to form an incoherent reservoir of excitons that collect around the polariton bottleneck region~\cite{Carusotto_RMP2013}. The reservoir not only provides gain for polaritons but also locally blueshifts the polariton due to strong exchange interactions between excitons and polaritons, effectively forming an optical trap~\cite{Askitopoulos_PRB2015}. At sufficiently high excitation densities (P~$\geq$~P$_{th}$), bosonic stimulation into a given trap state is triggered with subsequent formation of the nonequilibrium polariton condensate in the trap center. Under continuous wave excitation, the condensate losses (in the form of cavity PL) are balanced through the continuous supply of new particles from the reservoir.

The precise control of the confinement and position of the condensates in the plane of the cavity ($x$-$y$ plane) allows for the modification and adjustment of the polaritons' energy levels, enabling the construction of an optical analogue of a macroscopic 2D molecule and the study of interactions (spatial coupling) between condensates in two or more optical traps. Due to their optical nature, polariton condensates can be measured directly through the cavity PL, which gives direct information on the spatial amplitude and phase of the macroscopic wavefunction. 

\begin{figure}[t]
\centering
\includegraphics[width=1\linewidth]{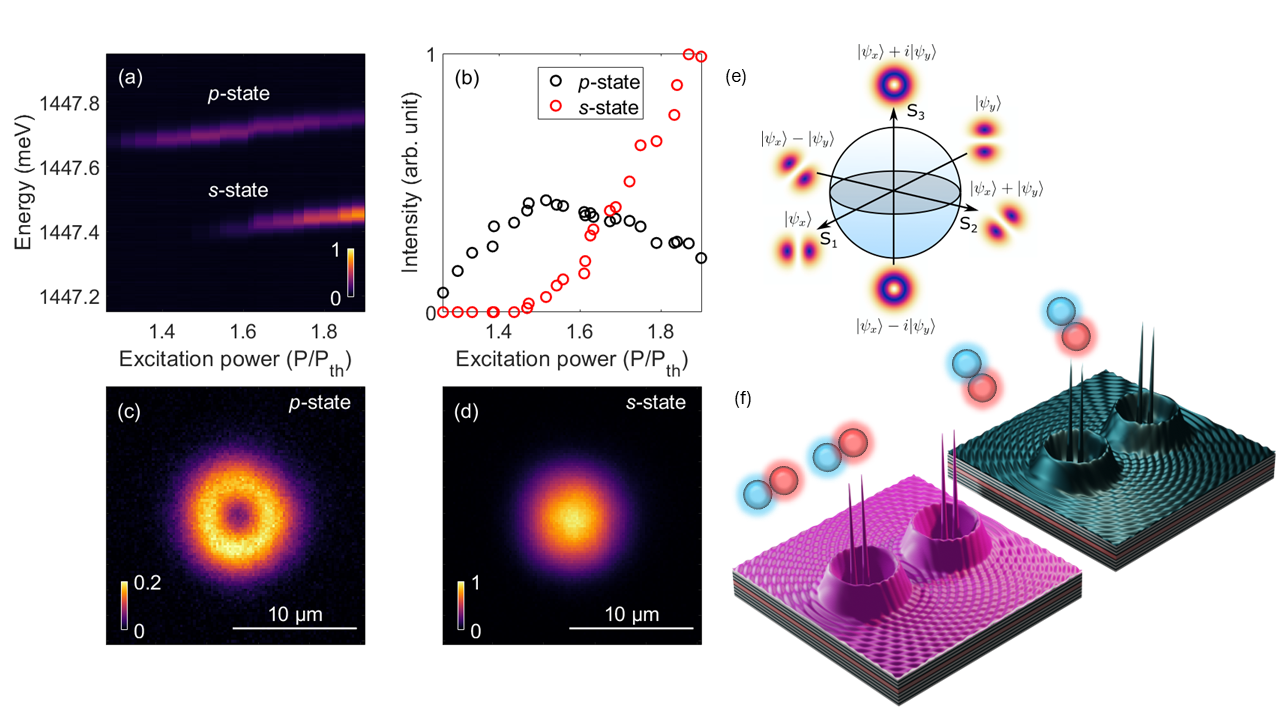}
\caption{Interaction-induced $\sigma$-$\pi$ crossover of the bonding configuration in trapped coupled polariton condensates. (a) Experimentally measured spectra of single, trapped condensate emission as a function of excitation power. (b) Integrated emission of $s$- and $p$-states. Total real-space photoluminescence (c) $< 1.6 P_{th}$ ($p$-state dominant) and (d) $> 1.6 P_{th}$ ($s$-state dominant). (e) Bloch sphere representations of the two-level $p$-state. (f) Schematic illustration of the investigated phenomena. The ring-shaped optical traps are used to create two spatially separated and ballistically coupled condensates, each residing in its own trap. The $p$-states alignment depends on the separation distance $d$ between traps. Red and blue spheres schematically represent the phase of each condensate dipole state illustrating in-phase (pink) $\sigma$-bonded and (green) $\pi$-bonded configurations between the traps.}
\label{fig1}
\end{figure}

While polaritons generally condense into the ground state at the bottom of the optical trap, their nonequilibrium nature permits them to condense into higher-order modes should the balance between gain and losses become optimal in such states~\cite{Cristofolini_PRL2013, Dreismann_PNAS2014, Askitopoulos_PRB2015, Sun_PRB2018, Topfer_PRB2020, Sitnik_PRL2022, Pieczarka_OptExpr2022, Aladinskaia_PRB2023, Barrat_SciRep2024, Zaremba_AdvOptMat2025}. At relatively low pumping powers (less than 1.6~P$_{th}$), the condensate is found to dominantly occupy the first excited state ($p$-state) manifold of the trap as shown in the spectrally resolved PL in Figs.~\ref{fig1}(a,b). At higher powers where energy relaxation of polaritons becomes more efficient, the ground state ($s$-state) of the trap dominates~\cite{Topfer_PRB2020} as seen from the red curve overtaking the black in Fig.~\ref{fig1}(b). A typical PL coming from the $p$-state below threshold is shown in Fig.~\ref{fig1}(c) and $s$-state above threshold in Fig.~\ref{fig1}(d). From here on, we will restrict our experiment to pump powers where the ground state is negligibly populated by the condensate and the physics of two coupled condensates is determined mainly by states in the $p$-manifold of the traps.

For a cylindrically symmetric trap, the $p$-state manifold forms a degenerate two-level quantum system which can be described in the basis of $|p_x\rangle \equiv |\psi_x\rangle$ and $|p_y\rangle \equiv |\psi_y \rangle$ states. These are dipole-shaped wavefunctions oriented along the $ x$- and $ y$-directions of the trap. Since the condensate is restricted within the $p$-manifold, it becomes appropriate to describe the state of the condensate as a coordinate on the surface of a Bloch sphere shown in Fig.~\ref{fig1}(e) with an associated Bloch vector,
\begin{equation}
\mathbf{S} = \begin{pmatrix} S_1 \\ S_2 \\ S_3 \end{pmatrix} = \boldsymbol{\Psi}^\dagger \boldsymbol{\hat{\sigma}} \boldsymbol{\Psi}.
\end{equation}
Here, $\boldsymbol{\hat{\sigma}}$ is the Pauli matrix vector, and the condensate is assigned an order parameter in the form of a macroscopically coherent wave function $\psi(\mathbf{r},t)$. Projecting the spatial order parameter onto the $p$-manifold we can define an associated condensate spinor $\Psi = (\psi_+,\psi_-)^\text{T}$ where $|\psi_\pm \rangle = (|\psi_{x} \rangle \pm i | \psi_{y} \rangle )/\sqrt{2}$. It is also useful to keep a separate lowercase notation for the normalized Bloch vector components $\mathbf{s} = \mathbf{S}/S_0 = \mathbf{S}/( \boldsymbol{\Psi}^\dagger \boldsymbol{\Psi})$ where $S_0 = |\psi_+|^2 + |\psi_-|^2$. For example, a $|p_x\rangle$ state would have $s_1 = 1$, and a $|p_y\rangle$ state $s_1=-1$. Condensate dipole states oriented along the diagonal and the antidiagonal are $s_2 = \pm 2$ and counterclockwise and clockwise vortices are $s_3 = \pm 1$ [see Fig.~\ref{fig1}(e)]. 

While the ladder of polariton condensate energy modes in a single optical trap has already been well studied~\cite{Pieczarka_PRB2019}, only recently was the directional coupling mechanism between two $p$-state polariton condensates investigated as a function of separation distance~\cite{Cherotchenko:PRB2021}. The dipole-shaped structure of the $p$-state condensates results in a unique interference pattern, as illustrated schematically in Fig.~\ref{fig1}(f), depending on whether the condensate dipole axis is along the axis connecting the traps or orthogonal to it. Due to their driven-dissipative nature, the condensates seek to interfere constructively, much like coupled lasers competing over mutual gain~\cite{Parto_Nanophotonics2020}, and spontaneously synchronize into extended states or artificial polaritonic molecules~\cite{Galbiati_PRL2012, Zambon:PRA2020, Gianfrate_NatPhys2024}. Specific to this study, states of the coupled system that have mirror symmetry along the $x$- and/or $y$-axis can be classified into four categories as,
\begin{equation} \label{eq.states}
\begin{split} 
    \sigma\text{-bond}: \quad |\psi_\text{A(B)}\rangle &= (|\psi_{L,x}\rangle  \pm |\psi_{R,x}\rangle)/\sqrt{2}, \\
    \pi\text{-bond}: \quad |\psi_\text{C(D)}\rangle &= (|\psi_{L,y}\rangle  \pm |\psi_{R,y}\rangle)/\sqrt{2}. 
\end{split}
\end{equation}

Here, $L(R)$ refers to the left(right) condensate and $\pm$ denotes whether they are in-phase ($+$) or anti-phase ($-$). States $|\psi_\text{A} \rangle$ and $|\psi_\text{C}\rangle$ are shown in pink and green colormaps respectively in Fig.~\ref{fig1}(e). From a molecular chemistry perspective, such configurations can be labelled as a 2D polariton analogue of $\sigma$- and $\pi$-bonded orbitals.
\begin{figure}[htbp]
\begin{center}
\includegraphics[width=1\linewidth]{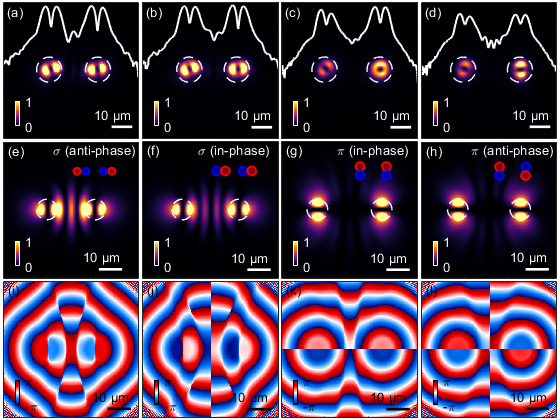}
\caption{Four possible coupling configurations of the $p$-states for a separation distance (a) $d = 22.9$ $\mu$m, (b) $d = 23.8$ $\mu$m, (c) $d= 26.7$ $\mu$m, (d) $d = 27.7$ $\mu$m measured as a real-space integrated emission. The continuous white line is a cross-section along the axis of interaction of the trapped condensates. The white dashed circles indicate the size of the traps. (e-h) Theoretically reproduced emission patterns corresponding to the configurations shown in panels (a-d). (i-l) Theoretically reproduced phase showing four possible types of coupling and alignment of dipole $p$-state condensates.}
\label{fig2}
\end{center}
\end{figure}

Figures~\ref{fig2}(a-d) show experimental results of the polariton condensate PL from two optical traps separated by different distances $d =$ 22.9, 23.8, 26.7 and 27.7 $\mu$m corresponding respectively to the states A, B, C, and D given by~\eqref{eq.states}. Applying mean field modeling, the same states can be found as fixed point solutions to the 2D generalized Gross-Pitaevskii equation at similar distances (see Methods). Figures~\ref{fig2}(e-h) show the simulated condensate density $|\psi(\mathbf{r})|^2$ and~\ref{fig2}(i-l) the phase $\arg{[\psi(\mathbf{r})]}$. High interference fringe contrast between the condensates evidences that polariton waves are emitted from each trap as propagating modes that lead to synchronization of the condensates. Such synchrony can also be found in so-called ballistic polariton condensates~\cite{Ohadi_PRX2016, Topfer_PRB2020} and in optically trapped polariton condensates populating the $s$-state manifold~\cite{Harrison:PRB2020}. The general behavior is that the separation distance $d$ is a crucial parameter in whether a given extended state will optimally constructively interfere and thus dominate over other possible states (i.e. take all the gain from the reservoir). We examined the stability of configurations A-D through extensive mean-field simulations (see section S2.A in Supplementary Information) by observing whether each configuration changes into another over time. We found that the stability of A and B $\sigma$-bonded configurations periodically alternates with distance $d$ with a little overlap, as expected~\cite{Cherotchenko:PRB2021}, and is otherwise stable in the lower range of pump powers (P$_{th}$ < P < 1.25 P$_{th}$). The stability of C and D $\pi$-bonded configurations, however, is heavily overlapped and only appears above a certain critical distance $d > d_{\text{crit},\pi}$.

A noteworthy difference between simulations and the experiment is the ``sharpness'' of the dipole structure in simulations. In the experiment, noise and disorder lead to smearing of the PL and possible triggering of the condensate into circulating currents, which make the condensate more annular rather than dipole-shaped~\cite{Askitopoulos_PRB2018}. We have theoretically studied this behaviour (see section S2.B in Supplementary Information) and found that the pump power is a crucial parameter in determining whether the condensate retains a sharp dipole-like form or becomes smeared into an annulus~\cite{Nalitov_PRA2019}. The former is dominant for small pump power, whereas the latter---the vortex state---becomes pronounced when the pump power is increased over a certain threshold value. It is worth noting that polariton condensates forming single-charge macroscopic vortices $|\psi_\pm \rangle$ in optical traps~\cite{Barrat_SciRep2024} have been realized using chiral static~\cite{Dall_PRL2014, Kwon_PRL2019}, rotating~\cite{Gnusov_SciAdv2023, Redondo_NanoLett2023}, and pulsed~\cite{Ma_NatComm2020, yao2025persistent} pumping schemes and recently explored as coupled elements in optical lattices~\cite{Alyatkin_SciAdv2024}.

\begin{figure}[htbp]
\begin{center}
\includegraphics[width=1\linewidth]{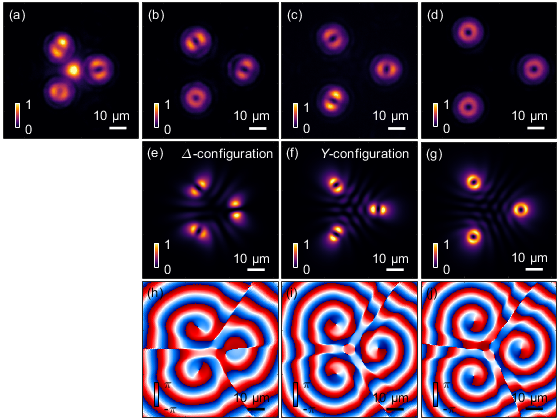}
\caption{Coupling between three equilaterally positioned optical traps containing $p$-state polariton condensates for varying separation distance (triangle edge length) (a) $d = 15.76$ $\mu$m, (b) 19.23 $\mu$m, (c) 22.17 $\mu$m, and (d) 26.85 $\mu$m. 
(a) When the traps are close together, the exciton reservoir creates an additional potential in the middle of the pattern, resulting in an additional bright emission spot at the center. When the pumps are further separated, the $p$-states of the trapped condensates orient themselves in (b) $\Delta$- and (c) Y-bonded configurations. (d) Above a critical distance when the interactions between traps are insufficient to affect the orientation of the {\it p}-states, they form in an annular shape. (e-g) Theoretically reconstructed $p$-state and (h-j) phase configurations corresponding to the experimentally observed configurations (b-d).}
\label{fig3}
\end{center}
\end{figure}

Figure~\ref{fig3} shows the case of three coupled trapped condensates in an equilateral triangle geometry for varying edge distances. In this case, a magnetic field of $B = 5$ T along the cavity growth direction (Faraday setup) is used to lower the condensation threshold by about $\approx 20\%$~\cite{Sawicki:PRB2024}. It allows us to obtain three condensates instead of two, using the same total excitation power, while keeping the system in the regime of a low probability of vorticity settling in each trap. We stress that the use of a magnetic field is not a necessary element of the experiment and is used here due to the unavailability of sufficient pumping power at 0T. It does not affect the orientation of the $p$-states since any effective spin orbit coupling of the polariton mode is quite weak in our sample~\cite{Cilibrizzi:APL2014}. For short distances shown in Fig.~\ref{fig3}(a) a unique scenario happens as the central space between the pumps effectively becomes a fourth optical trap simply because of the finite size and proximity of the other three annular beams at the vertices. This results in the formation of a fourth $s$-state condensate in the center of the triangle. While this is certainly an expected behavior~\cite{Cristofolini_PRL2013}, we are more concerned with the patterns appearing between the three $p$-state condensates at the vertices of the triangle in the absence of this central condensate, which otherwise blocks their coupling mechanism. We therefore focus our attention on Figs.~\ref{fig3}(b-d), which show the three coupled condensates for increasing edge distance, demonstrating spontaneous alignment in a $Y$-bonded setup [Fig.~\ref{fig3}(b)] with the poles aligned toward their centroid, and a $\Delta$-bonded setup [Fig.~\ref{fig3}(d)]. They are aligned perpendicular to the line between their centres and the centroid, and finally as in-phase vortices represented by an annular condensate density in each trap [Fig.~\ref{fig3}(d)]. In the lower panels of Fig.~\ref{fig3}(e-f) and~\ref{fig3}(h-j), we show the simulated condensate density and phase from mean field modeling (see Methods) reproducing the experimental patterns. We note that even a small projection on either pole of the Bloch sphere, implying finite vorticity in the trapped condensate, results in a spiral-like pattern in the condensate wavefunction phase (see section S2.B. in Supplementary Information).

\begin{figure}[htbp]
\begin{center}
\includegraphics[width=1\linewidth]{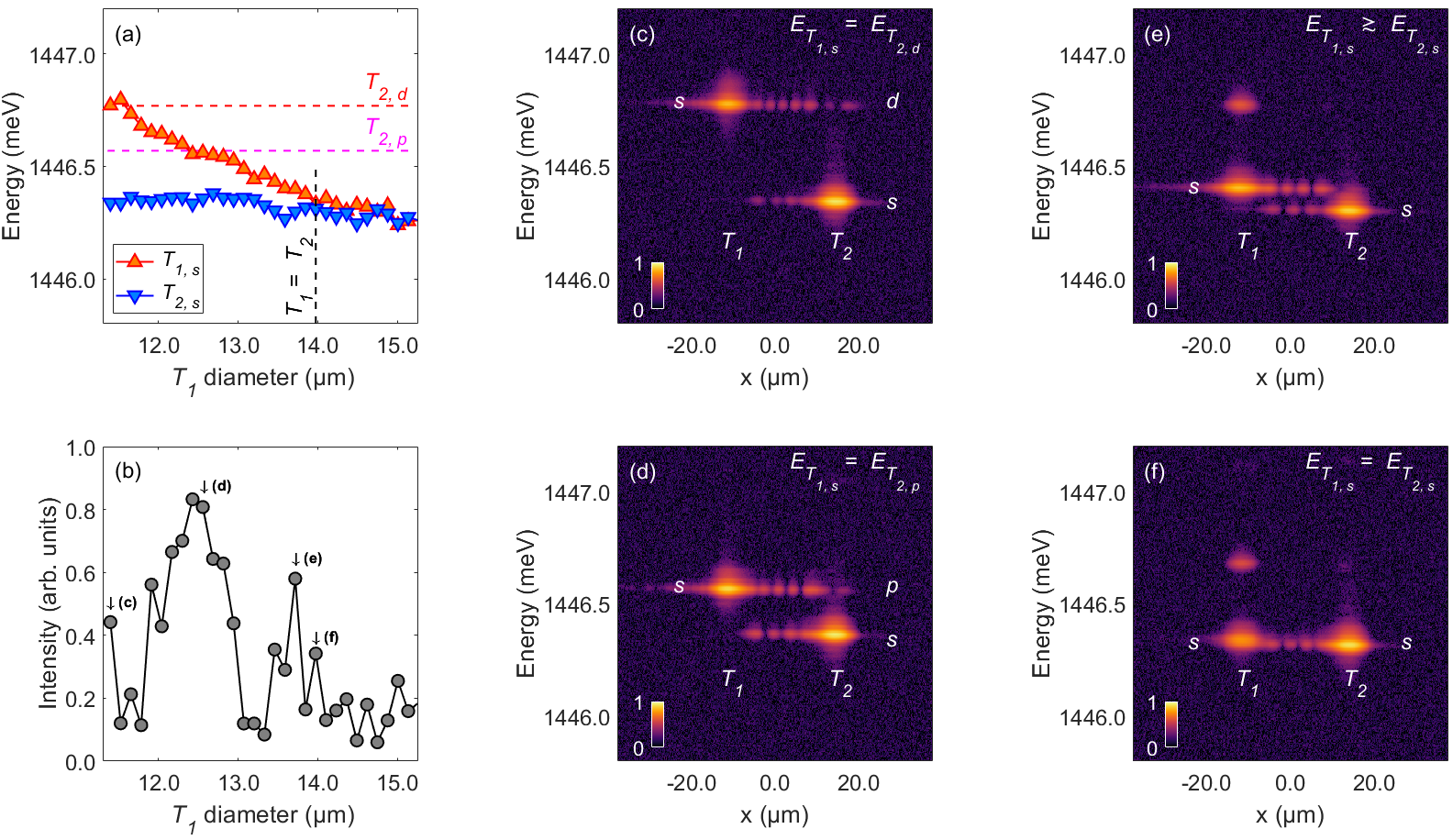}
\caption{Two coupled, trapped, mode-selectable polariton condensates. Independent control of the size of the trapping potential and thus the emission energy of one condensate relative to the other governs mode selection. (a) Energy of emission from {\it s}-states of coupled, trapped polariton condensates ($T_1$ and $T_2$). The orange triangles pointing up show the emission from $T_1$, which size was changed during the experiment. The blue triangles pointing down show the emission from $T_2$, which size was constant throughout the experiment. The magenta and red dashed horizontal lines show the energy of {\it p}- and {\it d}-states in $T_2$, respectively. The black vertical line indicates the situation with equal-sized traps. (b) Emission intensity of {\it s}-state of $T_1$ as a function of its size. When the emission energy is matched to the energy levels of $T_2$, the emission intensity increases. The letters in brackets refer to panels (c-f) of this figure, where the energy-resolved spatial distributions of the $T_1$ and $T_2$ trap modes are shown in the regimes of (c) $s$-$d$ coupling, (d) $s$-$p$ coupling, (e) detuned $s$-$s$ coupling with simultaneously occurring even and odd parity states, and (f) $s$-$s$ coupling. The imbalance in the population of {\it d}-states in $T_1$ and $T_2$ is attributed to the presence of disorder in the sample, which influences the spatial distribution of the emission.}
\label{fig4}
\end{center}
\end{figure}

In Figure~\ref{fig4}, we demonstrate the experimental mode selection by changing the relative size of the traps in a two-coupled trapped polariton condensate system. In the experiment, we show emission from spatially separated two trapped polariton condensates in which the size of one of the traps ($T_1$) is changed to achieve a coupling of the {\it s}-state of one trap to the selected mode of the other trap ($T_2$), while the distance between the condensates kept constant. We replace the annular optical traps with hexagonal traps composed of six laser spots placed in the corners of the hexagons (see section S3 in Supplementary Information). Polygons, similarly to the annular potential, cause polaritons to be trapped in the geometric centre of the potential created by the photoexcited excitonic reservoir~\cite{Cristofolini_PRL2013}. The use of hexagons instead of ring potential in this case has several advantages, including a reduction in the laser power required to create the condensate, easier control of the size of one of the traps and a cleaner pumping pattern. The use of hexagons also increases the losses of the trapping potentials, allowing for an increase in visibility of the interference fringes. Despite many advantages, polygon symmetry defines initial geometrical conditions that can determine the spatial alignment of $p$-modes. For this reason, to study subtle phenomena such as $\sigma$-$\pi$ switching, presented in the first part of this letter, we used circular traps, whereas in this part, we focus only on the spectral selection of the modes, neglecting the spatial pinning of modes. As before, to lower the condensation threshold, a constant magnetic field of $B = 5$ T is applied along the direction of cavity growth. 

Figure~\ref{fig4}(a) presents the energy of emission from {s}-states of both coupled, trapped polariton condensates as a function of $T_1$ size, while the size of $T_2$ remains unchanged. Changing the size of the $T_1$ modifies the energy of the trapped states populated by the condensate and also the spatial overlap between the condensate and the pumped exciton reservoir, which affects the pump-induced blueshift (i.e., optical nonlinearity)~\cite{Pieczarka_PRB2019}. By adjusting the size of one of the traps, one can then controllably match the energy of the fundamental mode ($s$-state) of one trap to any mode of the other trap with a fixed size. 

Figure~\ref{fig4}(b) presents the emission intensity obtained by fitting the Lorentz function to the emission spectra of $T_1$. The increases in intensity of the $T_1$ emission are observed when the energy of the {\it s}-state is tuned to the {\it s, p, d} modes of $T_2$ by changing the size of the $T_1$ trapping potential. The energy matching of modes between traps results in an increased constructive interference between the individual modes, leading to controlled asymmetric (between traps of different sizes) or symmetric (between traps of the same size) coupling between the two condensates. This is demonstrated by another property of the presented trapped artificial polariton molecules, which enables the optical system to mimic chemical systems where heteroatomic orbitals form bonds with orbitals of different azimuthal quantum numbers, such as $s$-$p$ or $s$-$d$. The subsequent panels Figure~\ref{fig4}(c-f) represent the spatially separated spectra of cases where the size of the $T_1$ is adjusted so that the corresponding {\it s}-state energy is matched to the selected $T_2$ modes. Figure~\ref{fig4}(c) shows the coupling of $s$- and $d$-modes obtained for trap diameters $D_{T{_1}}$ = 11.4 $\mu$m and $D_{T{_2}}$ = 14 $\mu$m, referring to the left and right trap respectively. Figure~\ref{fig4}(d) represents the $s$-$p$ coupling for traps of size $D_{T{_1}}$ = 12.6 $\mu$m and $D_{T{_2}}$ = 14 $\mu$m. The case Figure~\ref{fig4}(e) shows that when the traps are slightly detuned ($D_{T{_1}}$ = 13.7 $\mu$m and $D_{T{_2}}$ = 14 $\mu$m), two modes arise which couple to different parities - even and odd parity states. The symmetric case ($D_{T{_1}}$ = 14 $\mu$m and $D_{T{_2}}$ = 14 $\mu$m) Figure~\ref{fig4}(f) shows $s$-$s$ coupling.

\section{Conclusion}
Polaritonics has attracted interest as a possible platform for simulation of molecular systems, mapping molecular orbitals to coupled trapped polariton condensates~\cite{Galbiati_PRL2012, Gianfrate_NatPhys2024}. These systems have a high level of reconfigurability and tunability, which allows the system to be adapted to the problem under study.
While previous studies focused on condensates in their non-degenerate center-of-mass ground state, we were able to extend this approach to condensates in excited states, thus allowing us to simulate binding between multiple atomic orbitals. The presence of non-s orbitals leads to the possibility of different molecular geometry realisations with $\sigma$ or $\pi$ bonding. Using two polariton condensates, excited and trapped using SLM-generated patterns, we demonstrate that adjusting the distance between the traps enables the system to switch from a $\sigma$ to a $\pi$ configuration. 
 We also showed that in the case of three traps placed at the same distance from each other in a triangular geometry, a similar switching occurs (from Y- to $\Delta$-bond configuration). These changes were simulated with good agreement based on a mean field model. Moreover, the presented platform offers a reconfigurable geometry of the optical trap, and consequently, mode selection, which can be applied to control the type and strength of interaction in a complex optical system with discrete polaritonic states. Our work demonstrates an approach that provides a high degree of reconfigurability, which may find applications in various fields, such as optical switches or polariton lattices with intentionally introduced disorder.~\cite{Ohadi:PRB2018} 
Moreover, our platform, with the support of machine learning techniques such as the Fourier Neural Operator~\cite{Wang:arxiv2025}, which has recently been applied to recognise emission patterns from coupled polariton systems, constitutes a step forward in simulating molecular systems.

\section{Materials and Methods}
\subsection{Experimental details}
The sample used in these experiments is a high-quality ($Q\sim12000$), strain-compensated 2 $\lambda$ GaAs microcavity grown with molecular beam epitaxy.~\cite{Cilibrizzi:APL2014} It contains three pairs of 6 nm In$_{0.08}$Ga$_{0.92}$As QWs placed at the central three antinodes of the electric field and two QWs located at the two extreme anti-nodes of the microcavity. The front and back mirrors consist of GaAs and AlAs$_{0.98}$P$_{0.02}$ layers that formed 23 and 26 Bragg pairs, respectively. The experiments are performed at a negative exciton-cavity mode detuning of approximately $-2$ meV.

The sample is kept at a temperature of 7K in a continuous-flow liquid helium cryostat operating as a closed-cycle system and equipped with a superconducting coil that generates a magnetic field of up to 5 T. The polariton condensation is obtained under non-resonant continuous-wave pumping conditions using a single-mode circularly polarised Ti:sapphire laser tuned to the Bragg reflector's minimum ($\lambda_{\mathrm{exc}}=758.8\,\mathrm{nm}$). A spatial light modulator shapes the laser beam into one of four variants of stimulation: single ring, two identical rings separated spatially, three identical rings, or two hexagonal polygons of different sizes. For two and three traps, the laser power is set to 90 mW. To provide sufficient excitation power for three traps, the lasing threshold is lowered using a 5T magnetic field. To avoid heating effects in the sample, the laser is chopped at a frequency of 10 kHz with a duty cycle of 5\% with an acoustic optical modulator. 

\subsection{Generalized Gross-Pitaevskii simulations}

The scalar dynamics of the lower branch of an exciton-polariton condensate optically pumped by a nonresonant CW pump-profile $P(\mathbf{r})$ can be described in the mean field approximation, resulting in a generalised Gross-Pitaevskii equation for the condensate wave-function, $\psi(\mathbf{r}, t)$, coupled to an exciton reservoir density, $n_R(\mathbf{r}, t)$.
\begin{align}
i\partial_t \psi &= \left[-\frac{\hbar}{2m}\nabla^2 + \alpha |\psi|^2 + G\left(n_R + \frac{\eta P}{\Gamma}\right) + \frac{i(Rn_R - \gamma_{LP})}{2}\right]\psi\label{gp} \\
    \partial_t n_R &= -(\Gamma + R|\psi|^2)n_R + P(\mathbf{r}) \label{nr}
\end{align}
Here, $m$ is the effective mass of the lower polariton branch, $\alpha$ and $G$ are repulsive interaction strengths between two polaritons in the condensate and a polariton with a reservoir exciton, respectively. $\Gamma$ is the exciton reservoir decay rate, $\gamma_{LP}$ is the decay rate of the lower polaritons, $R$ is the rate of stimulated scattering of polaritons into the condensate from the exciton reservoir, and $\eta$ is a phenomenological constant accounting for additional blueshift due to charge carriers and high-momentum exciton background. The pump profile is of the same form used in~\cite{Askitopoulos_PRB2015},
\begin{align}
    P(\mathbf{r}) = P_0  \frac{L_0^4}{(x^2 + \beta y^2 - r_0^2)^2 + L_0^4}
\end{align}
where $P_0$ is the pump strength and $L_0$ controls width of the pump ridges, and $r_0$ its radius, and $\beta$ for detrimental ellipticity in the pump geometry.

The method used here to simulate the condensate-reservoir system is a split-step Fourier method leveraging a GPU-accelerated fast Fourier transform algorithm. The parameters used are similar to those in previous studies on this sample~\cite{Topfer_PRB2020}: $m = \SI{0.32}{\milli\eV\pico\second\squared\per\micro\meter\squared}$, $\alpha = g_0|X|^4/N_\text{QW} = \SI{0.0004}{\micro\meter\squared\per\pico\second}$,  $G = 2 g_0/N_\text{QW} = \SI{0.002}{\micro\meter\squared\per\pico\second}$, $|X|^2 = 0.4$, $N_\text{QW} = 6$, $\eta = 2$, $\Gamma = \SI{0.1}{\per\pico\second}$, $R = \SI{0.016}{\micro\meter\squared\per\pico\second}$, and $\gamma_{LP} = \SI{0.2}{\per\pico\second}$. Here, $\hbar g_0 = \SI{10}{\micro\electronvolt\micro\meter\squared}$ is the 2D exciton-exciton interaction strength, $N_\text{QW}$ is the number of quantum wells in the sample, and $|X|^2 = 0.4$ is the exciton Hopfield fraction of the lower polariton under slight negative detuning.


\section{Acknowledgement}
This work was supported by the European Union Horizon 2020 program, through a Future and Emerging Technologies (FET) Open research and innovation action under Grant Agreement No. 964770 (TopoLight). K.S. and S.D.L. acknowledge the Leverhulme Trust, grant No. RPG-2022-037. H.S. acknowledges the project No. 2022/45/P/ST3/00467 co-funded by the Polish National Science Centre and the European Union Framework Programme for Research and Innovation Horizon 2020 under the Marie Skłodowska-Curie grant agreement No. 945339. V.K.D acknowledges the Icelandic Research Fund (Rann\'{i}s), grant No. 239552-051.

\bibliography{sample}

\end{document}


\maketitle

\section{Theory of a single trapped condensate} \label{single.condensate}

\subsection{Cylindrically symmetric case}

\noindent 
In this section, we show that for a single cylindrically symmetric optical trap, the polariton condensate has only two stable solutions corresponding to either clockwise or anticlockwise rotating vortex states~\cite{Alyatkin_SciAdv2024}. These states are $\psi_{1,0} \pm i \psi_{0,1}$ with respect to the Hermite-Gaussian basis projected onto the Bloch sphere in Fig.~1 in the main text. Later, we will show that dipole solutions $\psi_{1,0}$ and $\psi_{0,1}$ become stabilized when cylindrical symmetry is broken by making the trap elliptically shaped, thus lifting the degeneracy between $\psi_{1,0}$ and $\psi_{0,1}$.

We start by deriving a simpler form of the condensate equations of motion by projecting and truncating the dynamics of the 2D generalized Gross-Pitaevskii (2DGPE) onto the two-level $p$-manifold,
\begin{align}
i\partial_t \psi &= \left[-\frac{\hbar}{2m}\nabla^2 + \alpha |\psi|^2 + G\left(n_R + \frac{\eta P}{\Gamma}\right) + \frac{i(Rn_R - \gamma_{LP})}{2}\right]\psi\label{gp} \\
    \partial_t n_R &= -(\Gamma + R|\psi|^2)n_R + P(\mathbf{r}) \label{nr}
\end{align}
Here, we will assume for simplicity that the CW-driven excitonic reservoir follows its steady state solution adiabatically so that $\partial_t n_R = 0$, which gives,
\begin{equation}
n_R = \frac{P(\mathbf{r})}{\Gamma + R |\psi|^2} = \frac{P(\mathbf{r})}{\Gamma} \left(1 - \frac{ R |\psi|^2}{\Gamma} + \mathcal{O}{(|\Psi|^4)} \right).
\label{eq.X}
\end{equation}
In the last step, we have Taylor expanded the reservoir solution around small $R|\Psi|^2/\Gamma$, which is valid if the condensate is not pumped too far above the threshold. This allows us to write a simpler 2DGPE,
\begin{equation}
i  \frac{\partial \psi}{\partial t} = \left[ - \frac{\hbar }{2m} \nabla^2 + \alpha|\psi|^2 + \left(G + i \frac{R}{2}\right) \frac{P(\mathbf{r})}{\Gamma} \left(1 - \frac{R |\psi|^2}{\Gamma} \right) - \frac{i \gamma_{LP}}{2}  + G \frac{ \eta P}{\Gamma}\right] \psi 
\label{eq.2DGPE}
\end{equation}
We are interested in a condensate that occupies the trap's first excited $p$-state manifold, which can be written generally as a superposition of degenerate  clockwise and anticlockwise orbital angular momentum (OAM) states,
\begin{equation}
\psi(\mathbf{r},t) = \xi(r) \left( \psi_+(t) e^{i \theta} + \psi_-(t) e^{-i \theta} \right) e^{-i \omega t}.
\label{eq.basis}
\end{equation}
Here, $\mathbf{r} = (r,\theta)$ is the in-plane coordinate in polar form, $\psi_\pm(t) \in \mathbb{C}$ describe the phase and amplitude of each vortex component, $\omega$ is the energy of the condensate, and $\xi(r)$ is the radial steady state ($\partial_t |\Psi|^2 = 0$) profile of the condensate in a single optical trap. Plugging in this truncated basis into the 2DGPE and integrating out the real space dependence, exploiting the orthogonality of the states, we reduce our partial differential equation to only two coupled ordinary differential equations that describe the dynamics of each vortex component making up the condensate (up to an overall energy shift),
\begin{equation}
i \frac{d \psi_\pm}{d t}  = \left[i \tilde{p} + (\tilde{\alpha} - i \tilde{R}) (|\psi_\pm|^2 + 2 |\psi_\mp|^2)   \right] \psi_\pm,
\label{eq.cpm}
\end{equation}
We have set $\omega=0$ without loss of generality. The new coefficients are,
\begin{subequations}
\begin{align}	 \label{intp}
\tilde{p} & = \frac{R}{2 \Gamma} \int \xi(r)^2 P(r) \, d\mathbf{r} - \frac{ \gamma_{LP}}{2}, \\
\tilde{\alpha} & = \alpha \int \xi(r)^4 \, d\mathbf{r} -  \frac{GR}{\Gamma^2} \int \xi(r)^4 P(r) \, d\mathbf{r}, \\ \label{intR}
\tilde{R} & = \frac{R^2}{2 \Gamma^2} \int \xi(r)^4 P(r) \, d\mathbf{r}.
\end{align}
\end{subequations}
We will now work exclusively above the threshold, where $\tilde{p}>0$. We can scale our system through the transformations $t \to \tau / \tilde{p}$ and $\psi_\pm \to \psi_\pm  \sqrt{\frac{\tilde{p}}{\tilde{\alpha}}}$ which gives us a dimensionless form of the equations,
\begin{equation}
 \frac{d \psi_\pm}{d \tau}  = \left[1 - (i - \sigma) (|\psi_\pm|^2 + 2 |\psi_\mp|^2)   \right] \psi_\pm,
\label{eq.cpm}
\end{equation}
where $\tilde{R}/(\tilde{\alpha} \tilde{p}) = \sigma$. Equations~\eqref{eq.cpm} can be written in terms of the amplitude and phase of each mode $\psi_{\pm} = \sqrt{N_\pm} e^{i\phi_\pm}$,
\begin{subequations}
\begin{align}
\frac{dN_{\pm}}{dt} &=  2 \left( 1 - \sigma N_{\pm} - 2\sigma N_{\mp} \right) N_{\pm}, \label{Eq.SingleTrap_a} \\
\frac{d\phi_{\pm}}{dt} &=  N_{\pm} + 2  N_{\mp}. \label{Eq.SingleTrap_b}  
\end{align} 
\end{subequations}
We see that the change in the phase of the modes is trivially determined by the dynamics of their amplitudes. We therefore only need to focus on solutions of equation~\eqref{Eq.SingleTrap_a}
which has three equilibrium points,
\begin{equation*} 
\begin{aligned}[b]
&(\mathrm{I}) & &N_{+} =0, & &N_{-} =1/\sigma, \\
&(\mathrm{II}) & &N_{+} = 1/3\sigma, &  &N_{-} = 1/3\sigma, \\
&(\mathrm{III}) & &N_{+} =1/\sigma, &  &N_{-} =0. \label{Eq.EquilibriumPoints}
\end{aligned}
\end{equation*} 
It is easy to show that the only (and equally) stable equilibrium points are (I) and (III) through the eigenvalues of the Jacobian of equations~\eqref{Eq.SingleTrap_a}. If the eigenvalues $\lambda$ are positive, then the system is unstable, whereas the contrary implies stability. For solutions (I) and (III), the eigenvalues are degenerate and equal to $\lambda_{(I)} = \lambda_{(III)} = - 2  < 0$. For solution (II) the eigenvalues are $\lambda_{(II),1} = 2/3 > 0$ and $\lambda_{(II),2} = -2<0$ which forms a saddle point which is stable in one direction but unstable in the perpendicular direction. Thus, dipole states (II) are intrinsically unstable in a perfectly cylindrically symmetric system and the only stable solutions are pure vortices with 50/50 probability of clockwise (I) or counterclockwise (III) current forming determined by random initial conditions.
\begin{figure*}
\centering
\includegraphics[width=0.6\linewidth]{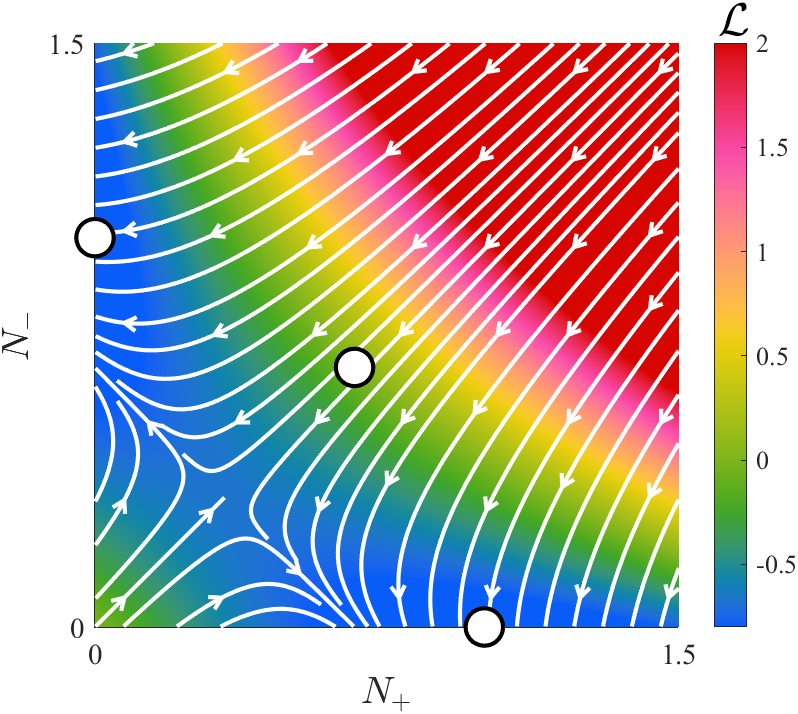}
\caption{The colorscale depicts the Lyapunov potential of the system with two minima (white dots) corresponding to clockwise and anticlockwise vortex solutions in which all phase space trajectories converge towards, underlining the robustness of the system. Here we set $\sigma = 1$.}
\label{flow}
\end{figure*}
The above argument can be better visulized by plotting the system Lyapunov potential with extrema corresponding to the solutions (I-III),
\begin{equation}
\mathcal{L} = -2(N_+ + N_-) +  \sigma(N_+^2 + N_-^2) + 4\sigma N_+ N_-
\end{equation}
The Lyapunov potential satisfies the condition $d \mathcal{L} /dt \leq 0$, which means that all points in phase space flow towards the two minima indicated by the white dots in Fig.~\ref{flow}. The two minima at the edges are the two counterrotating vortex solutions (I) and (III), and therefore any initial condition will always converge to either of these two points. The saddle point extrema are also clearly visible.

\subsection{Broken cylindrical symmetry}\label{broken.cylindrical.symmetry}
In the case of any inhomogeneities in either the sample location (disorder) or the pump profile (alignment effects causing the trap to be elliptically shaped), the system is no longer cylindrically symmetric, and the angular harmonics are no longer eigenmodes of the single-particle problem. The breaking of cylindrical symmetry can be described by a perturbative field $\epsilon \hat{\sigma}_x$ acting on the equations of motion, so they now read 
\begin{equation}
\frac{d \psi_\pm}{d t}  = \left[1  - (i + \sigma) (|\psi_\pm|^2 + 2 |\psi_\mp|^2)   \right] \psi_\pm - i \epsilon \psi_\mp.
\label{eq.cpm2}
\end{equation}
The choice of Pauli $\hat{\sigma}_x$ operator describes skewing (eccentricity) along the $x$-axis, which splits the energies of $p_x$ and $p_y$ dipole modes. The sign of $\epsilon$ dictates whether $p_x$ or $p_y$ is lower in energy. Here, we take $\epsilon>0$ and thus $p_y$ dipole is the single particle ground state. Intuitively, we expect that the condensate can become ``pinned'' into the $p_x$ state just like the linear polarisation pinning effect~\cite{Gnusov_PRB2020}, given the analogy between the vortex equations~\eqref{eq.cpm2} and polariton spinor equations of motion. However, one important difference is the ``cross-Kerr'' nonlinear term $2|\psi_\mp|^2$, which implies different behaviours above the condensation threshold in the nonlinear regime, since for polaritons, spins such a form of a term is typically negligible (i.e. exciton spin singlet interaction energy is very small compared to triplet interactions).
\begin{figure}[h!]
\centering
\includegraphics[width=1\linewidth]{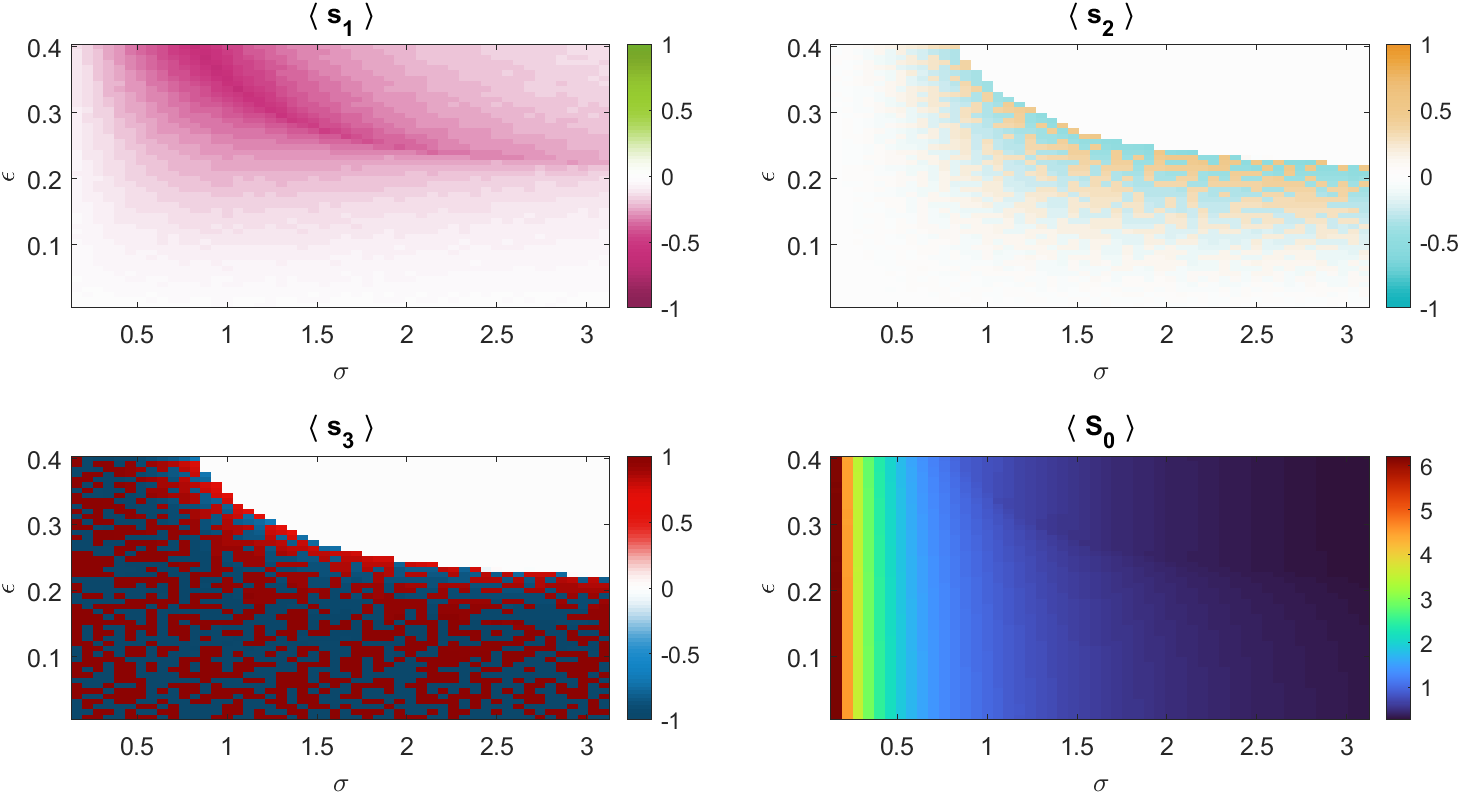}
\caption{Time average Bloch vector components from numerically solving~\eqref{eq.cpm2} for different values of $\sigma$ and $\epsilon$. Each pixel represents a random initial condition.}
\label{fig2}
\end{figure}

Figure~\ref{fig2} shows the time average normalized Bloch components $\langle s_i \rangle = 1/T \int_0^T \frac{S_i}{S_0} \,dt$ over a sufficiently long simulation time for different values of $\sigma$ and ``in-plane field'' strength $\epsilon$. Each pixel corresponds to a random initial condition. Two regimes can be observed, separated by a pink region corresponding to a dipole condensate ``pinned'' to the $p_y$ ground state. At low $\epsilon$ and $\sigma$ the condensate has a small $s_2$ projection and a strong $s_3$ projection corresponding to a dominant vortex regime. At high $\epsilon$ and $\sigma$, corresponding to large anisotropy and small condensate population, the Bloch vector precesses around the field $\epsilon \hat{\sigma}_x$ thus averaging out the $s_{2,3}$ components while retaining a slight projection on the $p_y$ ground state ($s_1<0$). When $\epsilon$ and $\sigma$ decrease, corresponding to lower anisotropy and stronger condensate population, the system bifurcates suddenly into a vortex state and obtains a large $s_3$ component and a slight $s_2$ component. The random red-blue pixelation in the $s_3$ component implies that the direction of the vortex rotation is randomly set by the stochastic initial conditions. This transition is markedly different and richer from the ideal cylindrically symmetric case, where only vortex states are found to be stable.

\begin{figure}
\centering
\includegraphics[width=0.99\linewidth]{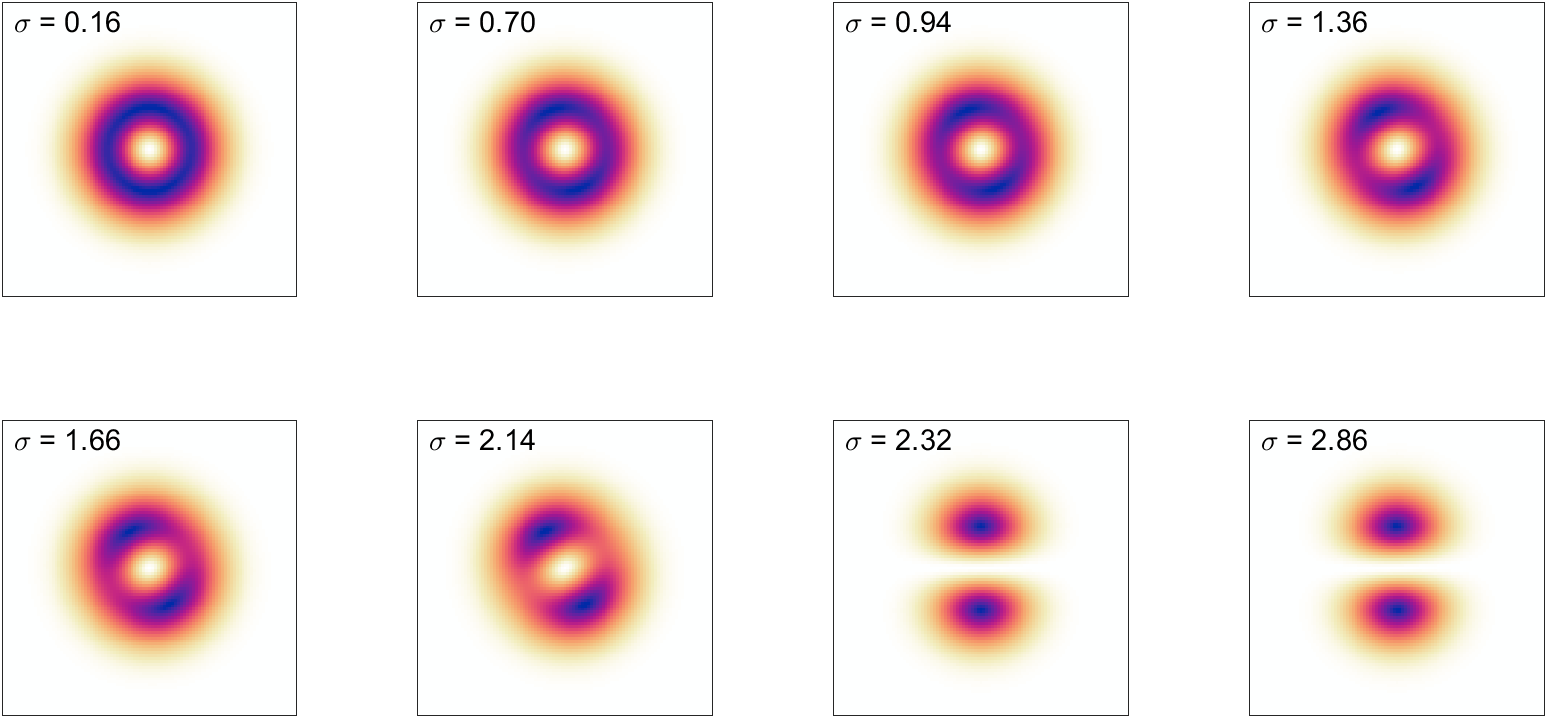}
\caption{Real space representation of the condensate density $|\psi|^2$ using the values of the average Bloch components from Fig.~\ref{fig2} for increasing values of $\sigma$ while fixing $\epsilon = 0.24$.}
\label{fig3}
\end{figure}

\noindent To give an idea of how the condensate emission would look like in time-integrated measurements, we plot in Fig.~\ref{fig3} our results projected back into the real space using the form,
\begin{equation}
\psi(\mathbf{r}) = \xi(r) \left(  \cos^2{(\Theta/2)} e^{i (\theta - \varphi/2)} + \sin^2{(\Theta/2)} e^{-i (\theta - \varphi/2)} \right).
\label{eq.basis2}
\end{equation} 
where $\Theta = \cos^{-1}{(s_3)}$ and $\varphi = \tan^{-1}{(s_2/s_1)}$ and $\xi(r) = r e^{-ar^2}$. The dipole transition towards a condensate vortex takes place around $\sigma \approx 2.2$.

\section{Multiple Trapped Condensates}

\subsection{Stability Analysis of Coupling Configurations}
When two optically trapped condensates are placed side by side they can couple stably in four ways: $\sigma$ bonded in/anti-phase and $\pi$ bonded in/anti-phase.

In simulations with random initial conditions for the order parameter, the condensates overwhelmingly favor forming in the $\sigma$ bonded configuration. However, this does not mean that the $\pi$ bonded configuration is unstable, and a $\pi$ bonded state can be produced with favourable initial conditions.

In order to explore the stability of each coupling regime, we simulate two trapped condensates over all combinations of 30 evenly spaced separations between the optical pump centres, $d$, going from \SI{12}{\micro\meter} to \SI{28}{\micro\meter}, and 30 evenly spaced pump powers, $P_0$, going from $P_{\text{th}}$ to $1.25P_{\text{th}}$. The threshold power is found for each $d$ by performing simulations with successively higher $P_0$, until the condensate intensity is found to increase or stay stable over time rather than decrease. We perform this simulation with four different initial conditions favourable to each coupling regime. Specifically, $\psi$ is initialised to a sum of two linear combinations of the first excited eigenstates of a 2D QHO with angular momentum $\pm 1$, $\phi_\pm$---i.e.~$\phi_{\pm}(\mathbf{r}) = \xi(r)e^{\pm i\theta}$---with origin at the center of their respective optical trap, which approximates $\sigma$/$\pi$ in/anti-phase coupling final states.

The system was time-evolved for 1 nanosecond. The traps had 10\% ellipticity along the y-axis, which enhanced the stability of $\pi$-bonded states. At regular intervals, including at the end of the simulation, Bloch parameters of the left and right condensates were calculated and recorded, as well as the phase difference between the condensates. Let $\psi_{L(R)}$ be the restriction of $\psi$ to the inside of the left (right) optical trap. The Bloch parameters were found by taking the projection of $\psi_{L(R)}$ onto $\phi_\pm$ to find the coefficients $\psi_{\pm,L(R)}$ such that $\psi_{+,L(R)}\phi_+ + \psi_{-,L(R)}\phi_-$ is the closest approximation of $\psi_{L(R)}$ that can be formed by linear combinations of $\phi_\pm$. Defining $\vec{\psi}_{L(R)} = [\psi_+, \psi_-]$, the Bloch components $S_{i,L(R)}$, where $i\in\{1,2,3\}$ are then calculated as
\begin{equation}
    S_{i,L(R)} = \frac{\vec{\psi}_{L(R)}\cdot\sigma_i \vec{\psi}_{L(R)}}{\vec{\psi}^2}
\end{equation}
where $\sigma_i$ are the Pauli matrices.

The phase difference between the condensates is found by taking the argument of the integral, $\Delta\phi = \arg\left(\int\psi_L^*(x,y)\psi_R(x,y)dxdy\right)$, where the right condensate is shifted so that its centre coincides with the left condensate's. The polariton number $N$ is calculated as $\int|\psi|^2dxdy$.

The final values of the Bloch components, phase difference and polariton number are summarised in Fig.~\ref{fig:initialconditiona}. 

\begin{figure}[h!]

    \centering
\begin{subfigure}[h!]{0.5\textwidth}
    \includegraphics[width=\textwidth]{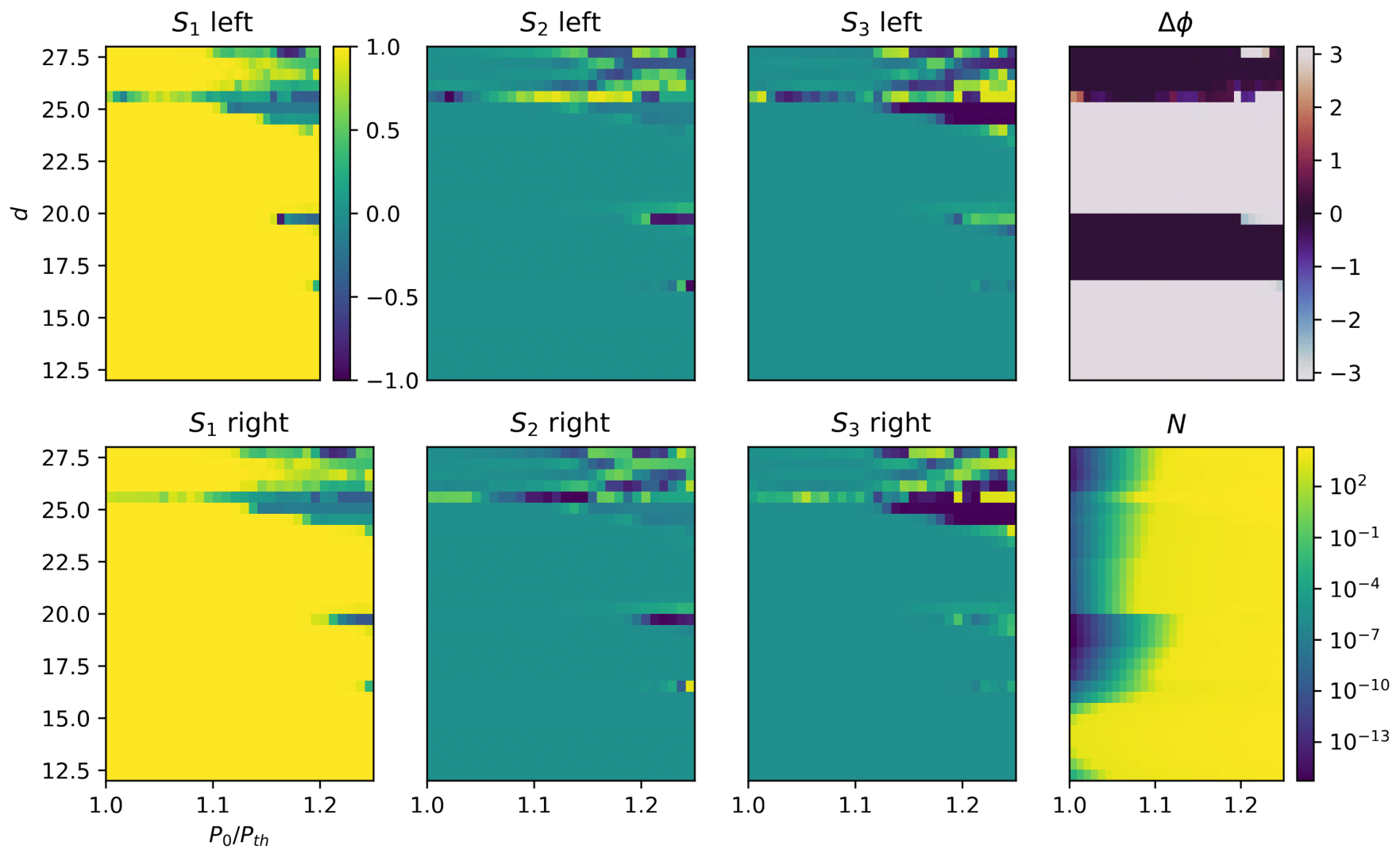}
    \caption{$\sigma$ bonded in-phase initial condition.}
    \label{fig:initialconditiona}
\end{subfigure}%
\begin{subfigure}[h!]{0.5\textwidth}
    \includegraphics[width=\textwidth]{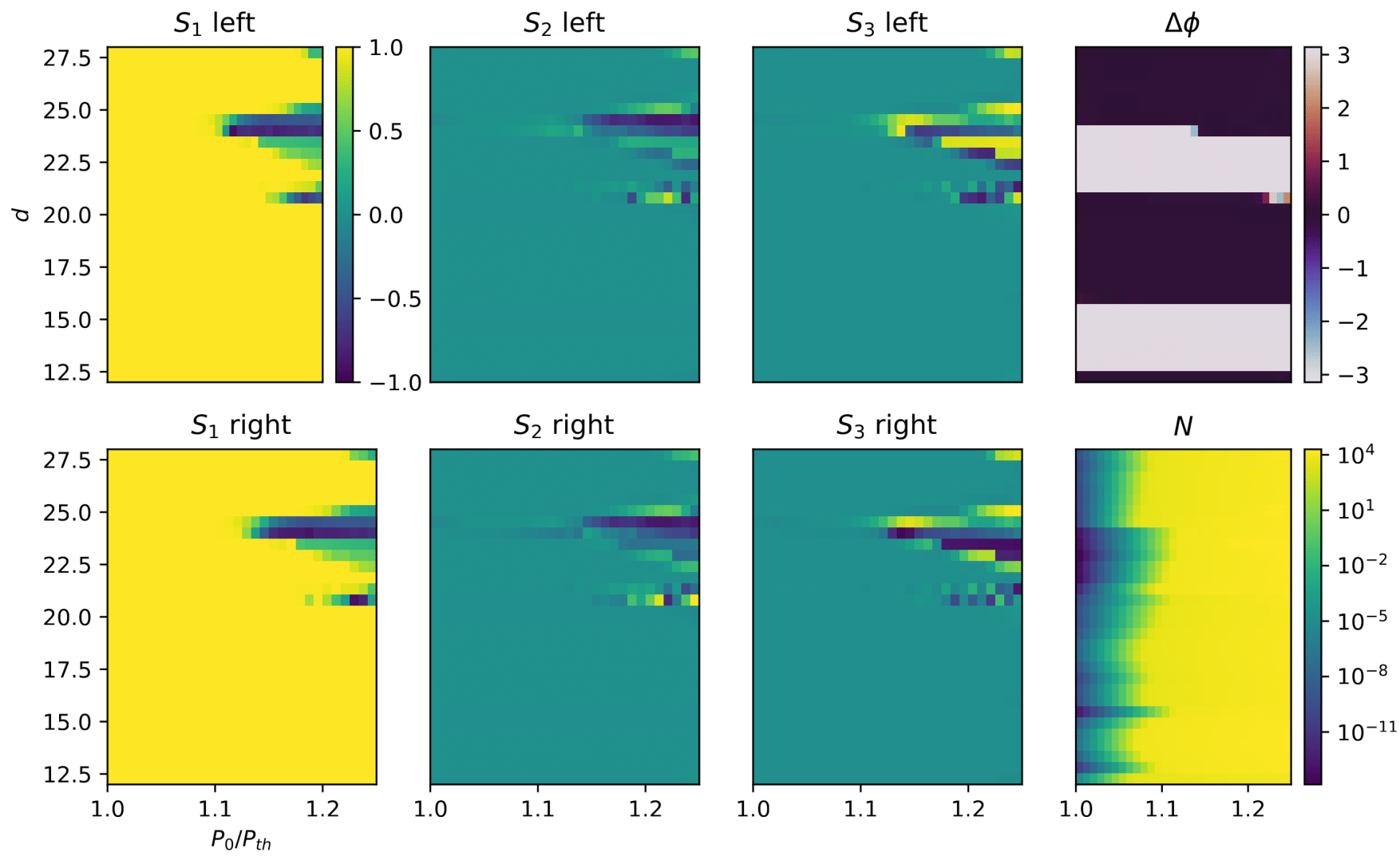}
    \caption{$\sigma$ bonded anti-phase initial condition.}
    
\end{subfigure}

\begin{subfigure}[h!]{0.5\textwidth}
    \includegraphics[width=\textwidth]{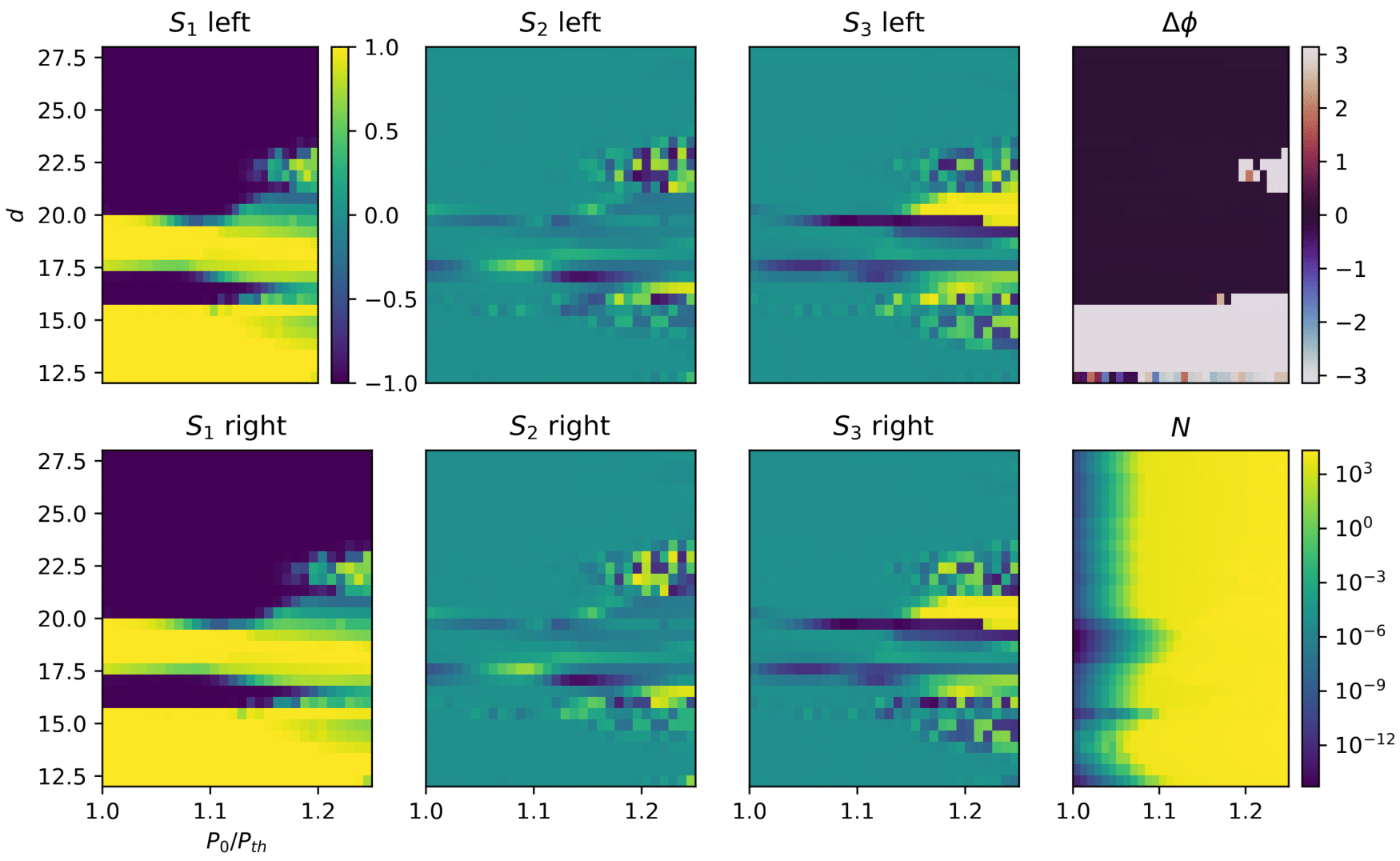}
    \caption{$\pi$ bonded in-phase initial condition.}
\end{subfigure}%
\begin{subfigure}[h!]{0.5\textwidth}
    \includegraphics[width=\textwidth]{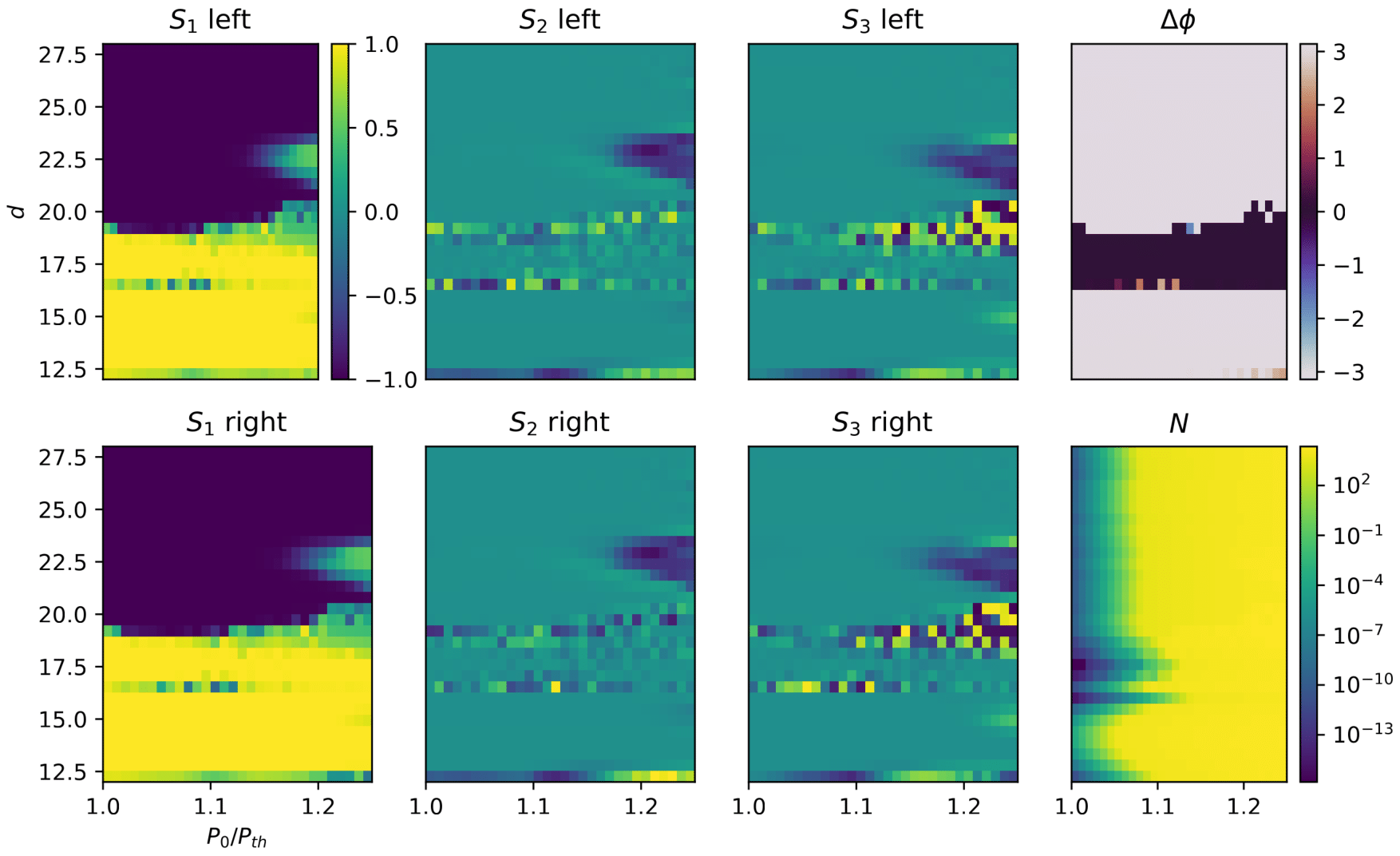}
    \caption{$\pi$ bonded anti-phase initial condition.}
\end{subfigure}
    \caption{Panels (a)-(d): Bloch components, phase differences, and polariton numbers of the simulation with various initial conditions.}
    \label{fig:initialconditiona}
\end{figure}

\begin{figure}[h!]
    \centering
    \includegraphics[width=0.9\linewidth]{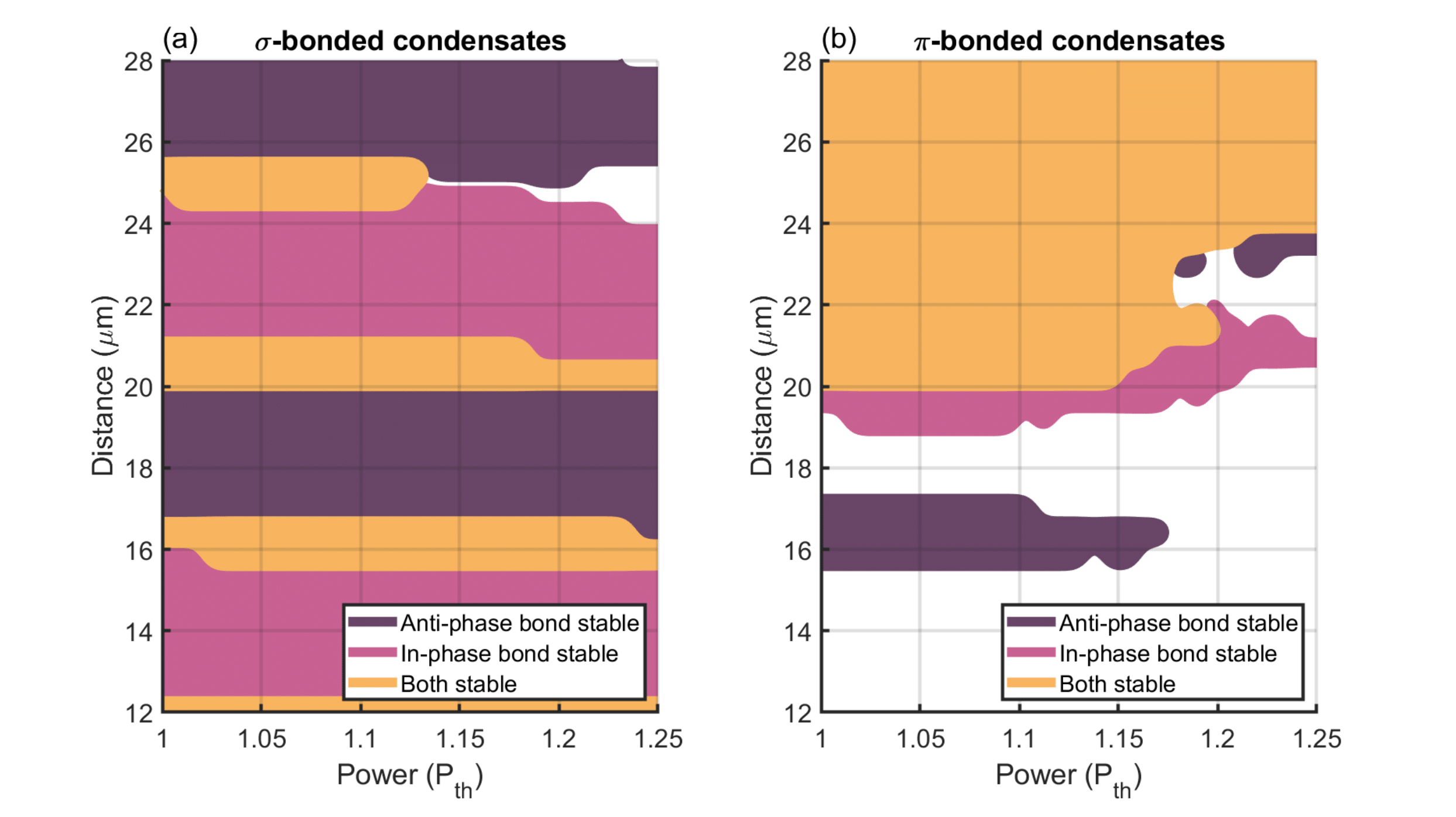}
    \caption{Colormaps showing for what combinations of pumping power and pump separation distance each kind of coupling is stable.}
    \label{fig:phasemap}
\end{figure}

Figure~\ref{fig:phasemap} condenses the results shown in Fig.~\ref{fig:initialconditiona} into a phase map, showing the parameter space where each coupling regime is stable.

\subsection{Formation of Vortex States With High Pumping Strengths}
\begin{figure}[h!]
    \centering
    \includegraphics[width=0.55\linewidth]{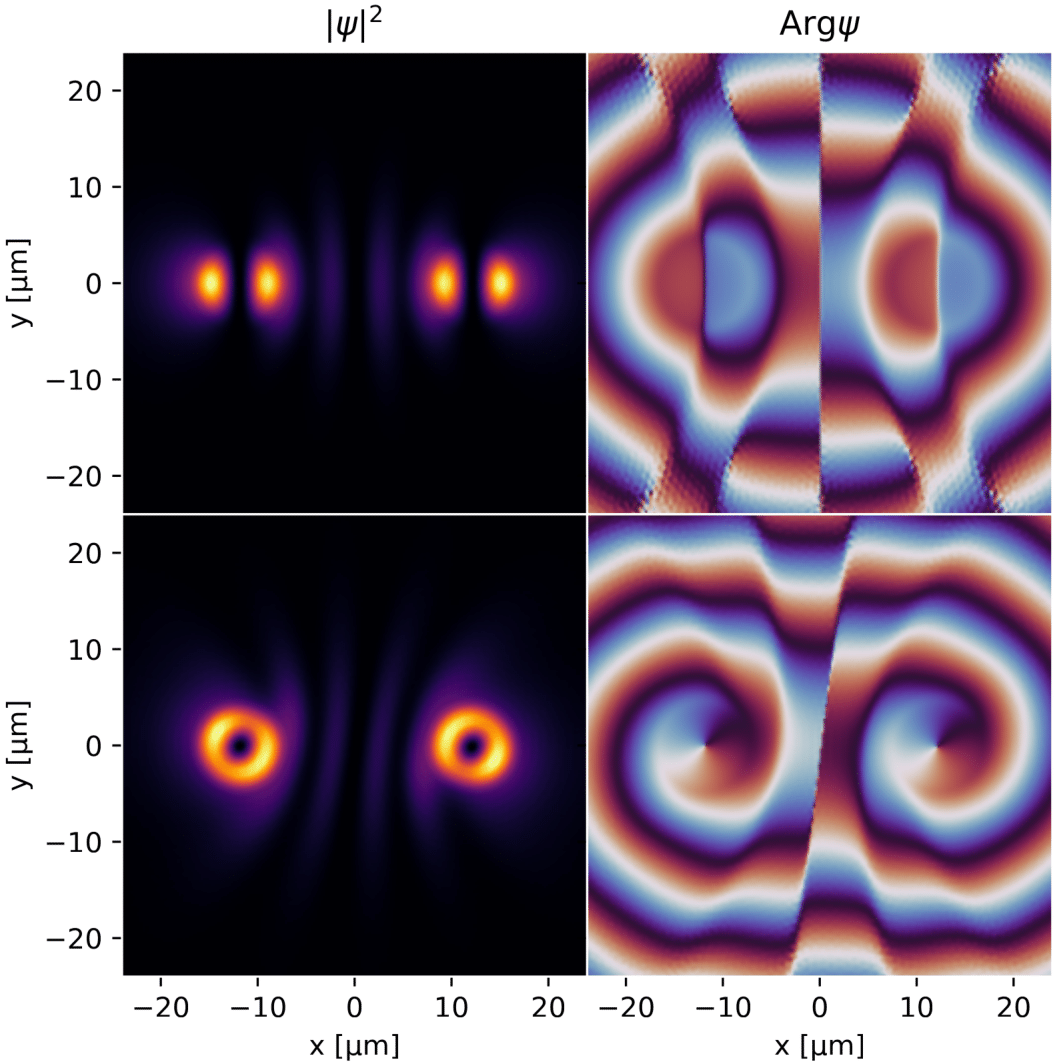}
    \caption{\textbf{Top:} Intensity and phase map of a numerically simulated order parameter starting from white noise initial conditions at threshold pumping. \textbf{Bottom:} Intensity and phase map of numerically simulated order parameter with the same initial conditions as the top simulation at $P_0=1.818P_{\text{th}}$.}
    \label{fig:twovortices}
\end{figure}

With greater optical pumping, the influence of the pump on each condensate will be significantly greater than the intertrap interaction and cause the condensates to form in annular vortex states. This is analogous to the effect in subsection S1.B, where bipole states can be induced in an elliptical trap at low pump strengths. However, with increasing pump strength, these become unstable, and vortex states become stable. This effect is shown in figure~\ref{fig:twovortices}.
 
\newpage\section{Optical traps}
\subsection{Annular and hexagonal traps}
\begin{figure}[h!]
    \centering
    \includegraphics[width=1\linewidth]{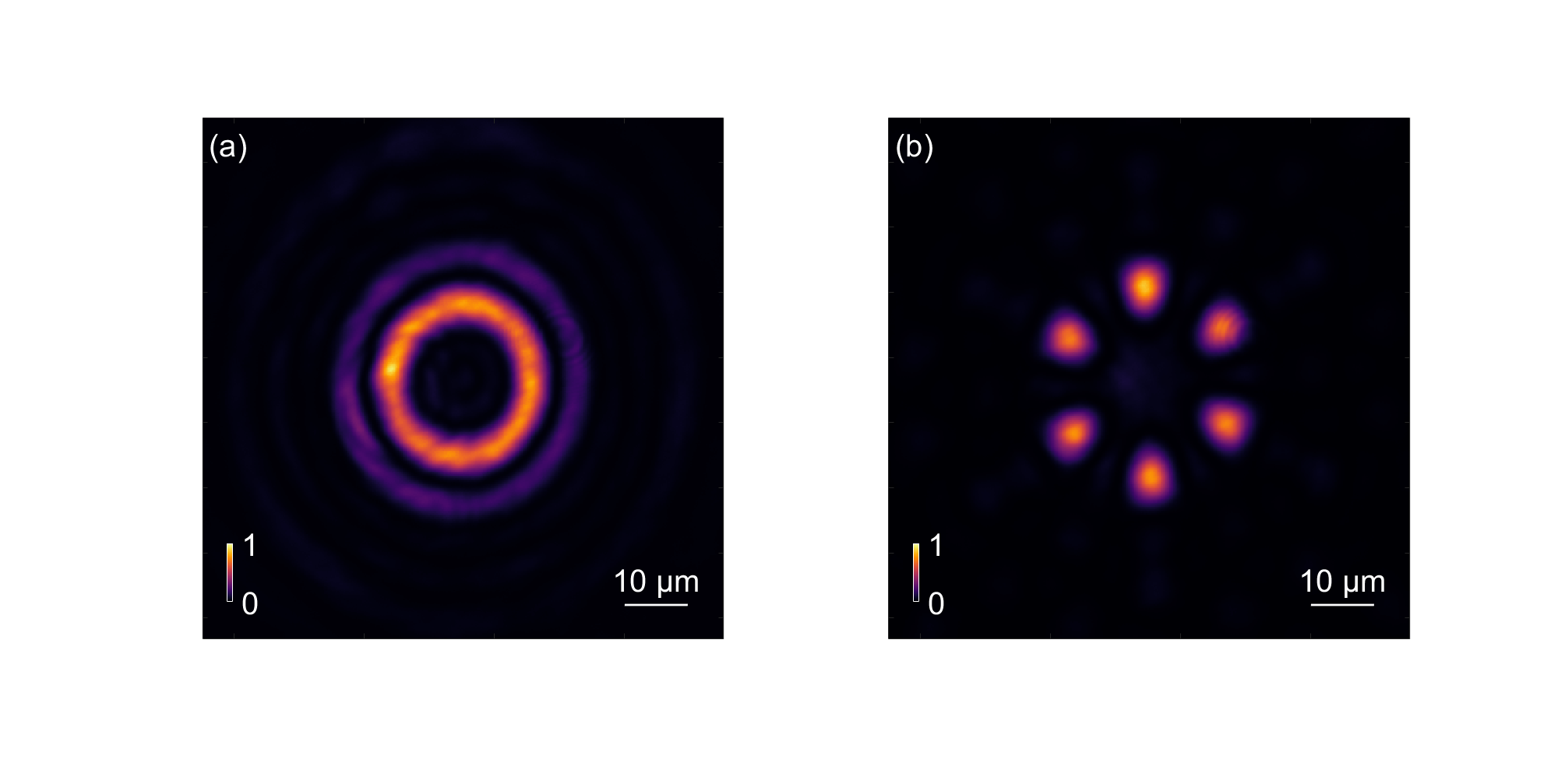}
    \caption{Two types of trapped potentials used in this work: (a) Annular trap, and (b) Hexagonal trap.}
    \label{fig:ringandhexagon}
\end{figure}

In this work, two types of optical traps with different optical properties were used. Figure~\ref{fig:ringandhexagon}(a) shows a laser profile formed in the shape of a ring trap. This type of trap provides high symmetry of the trapping potential, which does not favor any spatial orientation, which is why it was used to study the interactions of two and three identical traps. High symmetry enables the observation of $\sigma$-$\pi$ and $Y$-$\Delta$ switching. Figure~\ref{fig:ringandhexagon}(b) shows a hexagonal trap that is suitable for asymmetric coupling due to its more straightforward reconfigurability and the lower power required to form the trapping potential. 

\newpage
\bibliography{references}